\documentclass[aps,preprintnumbers,twocolumn,prl,showpacs,superscriptaddress]{revtex4-1}
\usepackage{amssymb}
\usepackage{amsmath}
\usepackage{amsfonts}
\usepackage{amsopn}
\usepackage{graphics}
\usepackage{graphicx}
\usepackage{subfigure}

\begin{document}

\title{Action diffusion and lifetimes of quasistationary states in the Hamiltonian Mean Field model}

\author{W. Ettoumi}
\author{M.-C. Firpo}
\affiliation{Laboratoire de Physique des Plasmas CNRS-Ecole
Polytechnique, 91128 Palaiseau cedex, France}

\date{\today}

\begin{abstract}

Out-of-equilibrium quasistationary states (QSSs) are one of the signatures of a broken ergodicity in long-range interacting systems. For the widely studied Hamiltonian Mean-Field model, the lifetime of some QSSs has been shown to diverge with the number $N$ of degrees of freedom with a puzzling~$N^{1.7}$ scaling law, contradicting the otherwise widespread $N$ scaling law. It is shown here that this peculiar scaling arises from the locality properties of the dynamics captured through the computation of the diffusion coefficient in terms of the action variable. The use of a mean first passage time approach proves to be successful in explaining the non-trivial scaling at stake here, and sheds some light on another case, where lifetimes diverging as $\mathrm{e}^N$ above some critical energy have been reported.

\end{abstract}


\maketitle

Explaining the emergence of quasistationary states~(QSSs), predicting their characteristics or determining their lifetimes are still puzzling issues in the active research program~\cite{Baldovin,Campa2008,Gabrielli2010,VdB,Chavanis2010,Gupta2011,Kastner2011} on long-range almost collisionless systems. Such systems are widely present in the Universe, since they range from assemblies of charged particles interacting via Coulomb interaction to self-gravitating massive objects such as globular clusters or stars in galaxies. Toy models have become a favorite tool to address those problems. For instance, the peculiar relaxation properties of long-range interacting systems began to be uncovered~\cite{Wright} through numerical simulations of the one-dimensional gravitational system, showing notably its reluctance to thermalize due to the existence of QSSs~\cite{Joyce}. Introducing periodic boundary conditions produced even simpler models permitting convenient computations in a compact space. Because it only retains the lowest Fourier mode of the gravitational potential, the well-known Hamiltonian Mean Field (HMF) model may be viewed as the simplest relevant toy model to address the intricate relationships between dynamics and statistical mechanics of long-range interacting systems. It is defined by the following Hamiltonian
\begin{equation}
\mathcal{H} = \sum\limits_{i=1}^{N}\frac{{p_i}^2}{2} +
\frac{1}{2N}\sum\limits_{i=1}^{N}\sum\limits_{j=1}^{N}\left[1-\cos\left(q_i-q_j\right)\right],
\label{eqn:hamiltonian}
\end{equation}
where $N$ is the number of particles, and $q_i$ and $p_i$ denote respectively the position and momentum of the $i^{\mathrm{th}}$ particle. A useful collective quantity to introduce is the mean-field (also called magnetization) vector $(M_x,M_y)$ with $M_{x} = 1/N \sum_i \cos q_i$ and $M_{y} = 1/N \sum_i \sin q_i$.
The average energy per particle $U=\mathcal{H}/N$ reads then
\begin{equation}
U = \sum\limits_{i=1}^{N}\frac{{p_i}^2}{2N} + \frac{1}{2}\left(1-M^2\right),
\label{eqn:U}
\end{equation}
where $M\equiv\sqrt{{M_x}^2+{M_y}^2}$ denotes the modulus of the magnetization vector. Equilibrium statistical mechanics~\cite{AntoniRuffo} can be rather easily derived and shows that a second order phase transition takes place at $U_{c}=3/4$. As $N$ tends to infinity, the ensemble average of the magnetization is accordingly positive for $U < U_{c}$ whether it is null for $U>U_{c}$.

However, contrary to short-range interacting systems, thermodynamic equilibrium may not be reached in the thermodynamic limit, which amounts, in the HMF model, to a Vlasov limit. Hence, QSSs are a signature of a \emph{broken ergodicity} \cite{Barre2001,Mukamel2005} in long-range interacting systems. Within the HMF model, their existence has been numerically demonstrated and discussed in various places (see e.g.~\cite{VLatora,Yamaguchi,Moyano,AntoniazziLB,Gupta2010,Pakter2011,Benetti2012,Rocha2012} and the recent review~\cite{Campa}). How finite-$N$ effects impact on the relaxation timescales towards equilibrium and on QSS's lifetimes in this mean-field model has been a central issue.

In this matter, leaving apart the far less studied case of QSSs in which the magnetization exhibits macroscopic oscillations~\cite{MoritaPRL}, two sorts of QSSs may be distinguished. Starting from some far-from-equilibrium initial distributions of $N$ particles, the system may reach a QSS having a lifetime $\tau$ diverging with $N$, associated either with a finite $O(1)$ or with a vanishing, namely $O(N^{-\beta})$ with $\beta > 0$, yet subcritical, magnetization. In both cases, the value of the magnetization in the QSS generically differs from its statistical ensemble average, although the discrepancy may be rather small, as e.g. in the initially monokinetic case \cite{EttFir1}. In the case of magnetized QSSs, i.e. with $M=O(1)$, numerical evidence and analytical arguments have supported the conclusion that $\tau$ should scale as $N$. However, the case of QSSs with a vanishing mean-field has given indications of another, nontrivial and still unexplained, scaling. Several numerical studies \cite{Yamaguchi,Zanette,CampaPRE} indicated that, in this case, the QSS lifetime $\tau$ should scale as $N^\alpha$ with $\alpha=1.7$ within some error bars. It is the prime objective of this Letter to provide for the first time an explanation of the latter intriguing scaling. Doing that, a more essential aim of this study is to illustrate a general phase-space approach to address the issue of QSS lifetimes.

The study proceeds along the following steps: After a brief account of the HMF QSSs phenomenology, the action-angle set of coordinates $(J,\theta)$ is introduced. After justifying a stochastic description, the local diffusion coefficient $D(J)$ is computed in the QSS regime. The QSS lifetime is shown to be controlled by the \emph{macroscopic fraction of the less diffusive particles}. A scaling argument predicts the diffusion coefficient in the corresponding action domain. Finally, this enters the computation of a mean first passage time, which serves as an indicator of the timescale of the action space visit and, consequently, of the thermalization timescale and QSS lifetime. This is shown to account for the $N^{1.7}$ scaling of the $M=o(1)$ QSS lifetimes. A discussion on the generality of the approach concludes the Letter.

\begin{figure*}[htbp]
\begin{center}
     \subfigure{\includegraphics[scale=0.5]{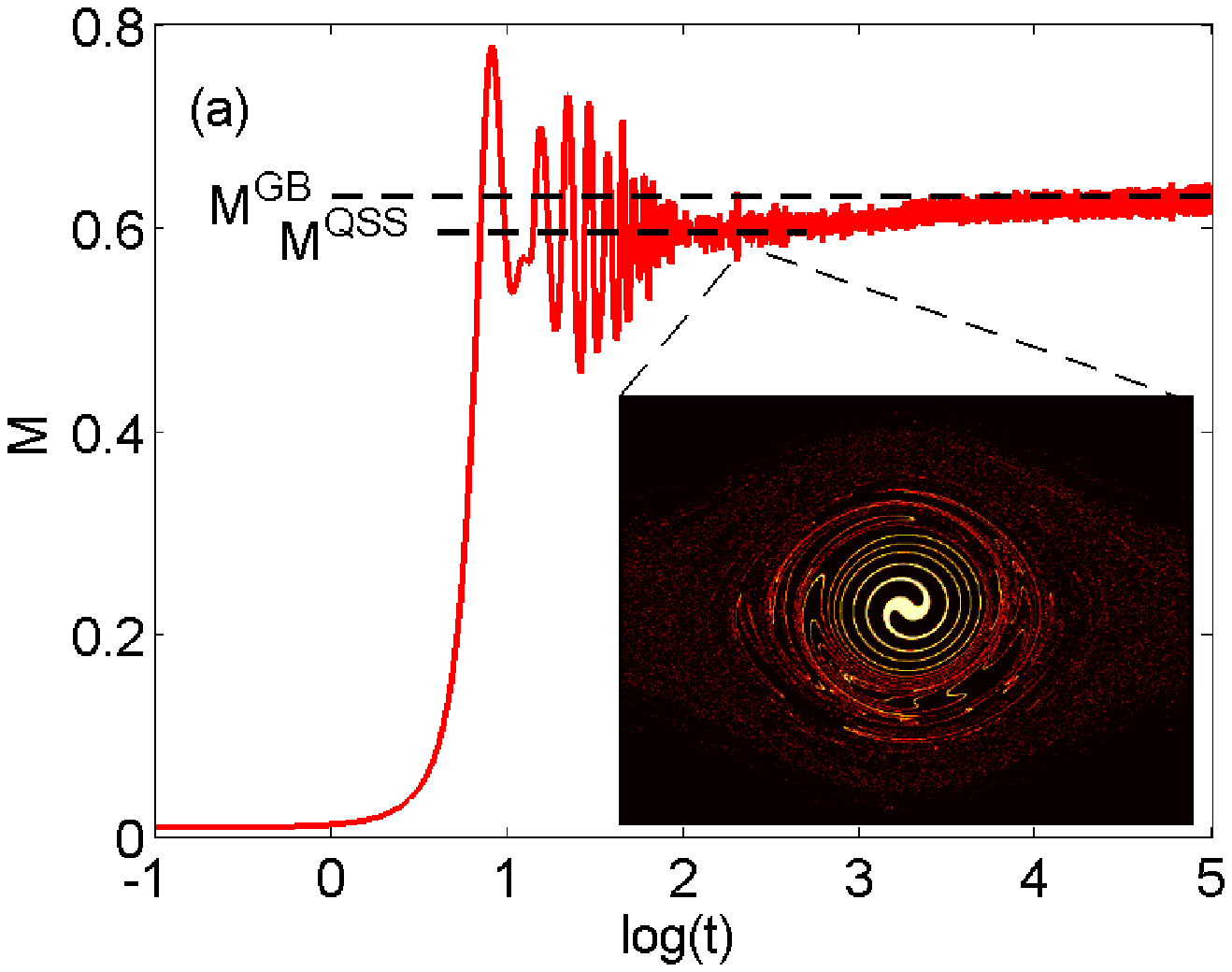}} \hfil \subfigure{\includegraphics[scale=0.5]{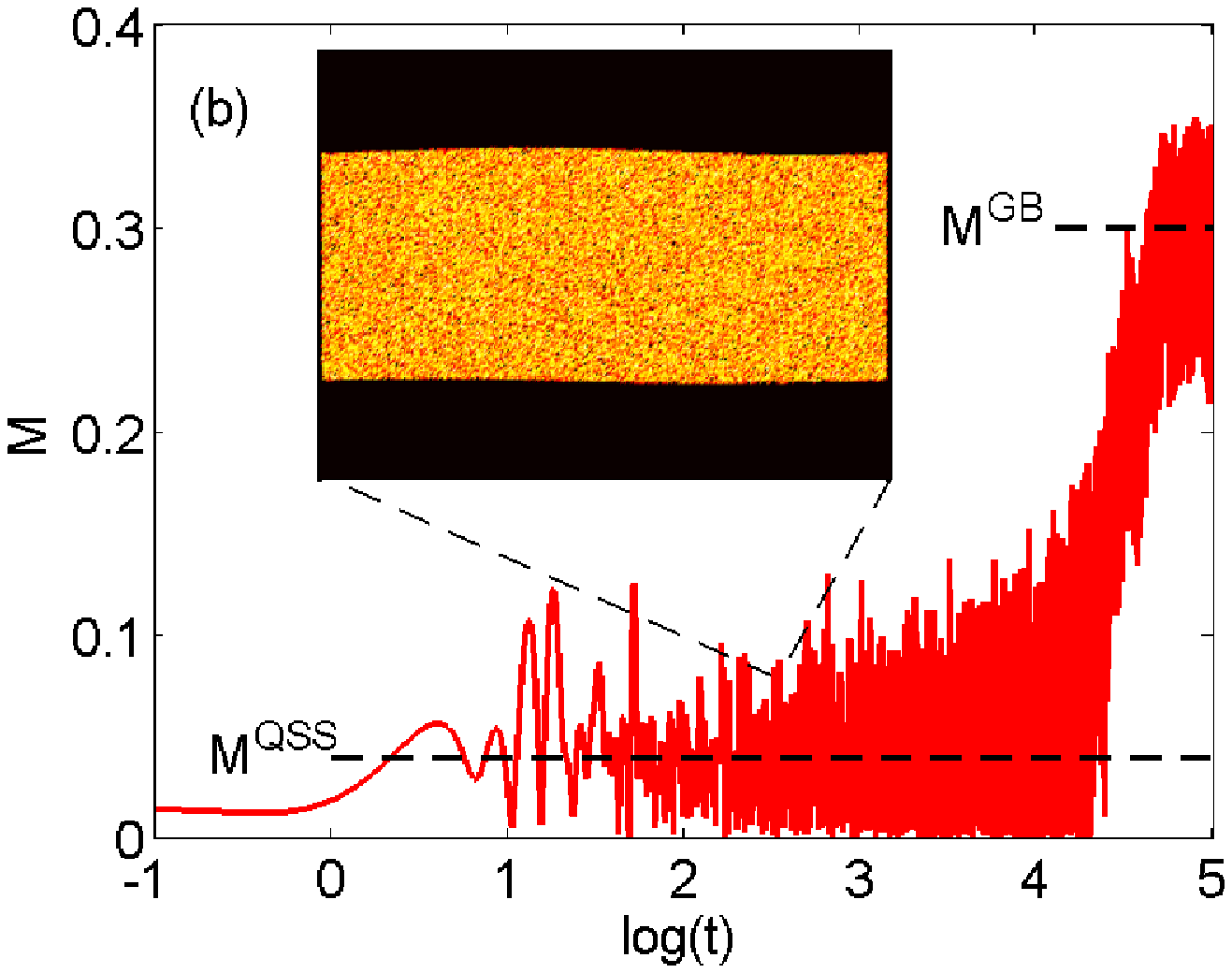}}
\end{center}
\caption{Time evolution of the mean field in log scale for two different initial conditions. (a) $N=10000$ particles are initially homogeneously spatially distributed and almost monokinetically, which corresponds to the Vlasov unstable waterbag case with $U\simeq 0.5$. This evolves through a violent relaxation towards a QSS which magnetization is quite close to the equilibrium one. In this case, the QSS lifetime is of order $N$. (b) Time evolution of a $U=0.69$ waterbag for $N=2000$, leading to a QSS in which the magnetization is of order $N^{-1/2}$. In this case, the QSS lifetime diverges as $N^{1.7}$~\cite{Yamaguchi,Zanette,CampaPRE}. The insets correspond to phase space snapshots during the QSS regime obtained with a much larger amount of particles for clarity purposes.}
\label{fig:evo_M}
\end{figure*}
Firstly, it is important to notice that the equation of motion of any particle $i$ in the HMF system (\ref{eqn:hamiltonian}) reads
\begin{equation}
\ddot{q}_i+M(t)\sin (q_i-\phi(t))=0,
\label{pendulum}
\end{equation}
where $M(t)$ and $\phi(t)$ are respectively the time-dependent modulus and phase of the mean-field. However, in the HMF model, the latter parameters are self-consistently determined by the motion of the $N$ pendula. From the definition of the mean-field and Eq.~(\ref{pendulum}), one can easily realize that the $M=O(1)$ QSSs correspond to metastable states in which a macroscopic, namely $O(1)$, fraction of particles are trapped in the mean-field potential trough. The system may reach such a state starting from Vlasov unstable initial conditions and passing through a violent relaxation phase \cite{EttFir1,EttFir2}, which is the situation depicted on Fig.~\ref{fig:evo_M}(a), or may get trapped into a magnetized Vlasov stable state. Contrarily, the $M=o(1)$ QSSs correspond to out-of-equilibrium configurations in which only an infinitesimal $o(1)$ fraction of particles are located into the mean-field potential well. In the literature, they have been obtained starting from Vlasov stable force-free out-of-equilibrium initial conditions, such as the so-called waterbag or semi-elliptical ones \cite{CampaPRE}. These are Vlasov stable equilibria having clearly non-thermal velocity tails. For the waterbag (WB) distribution function, $f_{\mathrm{WB}}(p,q)=\Theta(\Delta p -|p|)/(4 \pi \Delta p)$, where $\Theta$ stands for the Heaviside function, the condition for Vlasov stability is that $U\geq 7/12$. This corresponds to the situation shown in Figure \ref{fig:evo_M}(b) where $U=\Delta p^{2}/6+1/2=0.69$. The $M=o(1)$ QSSs may also be reached from magnetized unstable initial conditions \cite{Zanette}.

In order to investigate the issue of the divergence of QSSs lifetimes with $N$ in the $M=o(1)$ case and prepare the stochastic formulation that follows, it is convenient to decompose $M(t)$ into its QSS time-average $M_{0}$ and its fluctuating part $\delta M(t)$ and move to the action-angle variables $(J,\theta)$ associated to the non-perturbed, i.e. constant $M_{0}$, pendulum (\ref{pendulum}). The action $J$ is defined by $J=(2 \pi)^{-1} \oint p \,\mathrm{d}q$. In this framework, the instantaneous separatrices defined by $p_\mathrm{s}=\pm 2\sqrt{M_{0}} \cos( (q-\phi(t))/2))$ divide the phase space into rotational (passing) and libration (trapped) motions. Using for convenience a shifted action continuous at the separatrices and putting $J^{\ast}={8\sqrt{M_{0}}}/{\pi}$ the value of this action at the separatrices, and $h=p^2/2-M_{0}\cos(q-\phi(t))$, this is defined as
\begin{eqnarray}
\frac{J(h)}{J^{\ast}}=\left\{
\begin{array}{ll}
\mathrm{E}\left(k\right)-(1-k^2)\mathrm{K}\left(k\right)\text{ if } h\leq M_{0},\\[3mm]
k\mathrm{E}\left(k^{-1}\right)\text{ otherwise},
\end{array}
\right.
\label{eqn:Jh}
\end{eqnarray}
where we have used $k=\sqrt{1/2+h/(2M)}$, and where $\mathrm{K}$ and $\mathrm{E}$ denote respectively the complete Legendre elliptic integrals of first and second kind.

\begin{figure}[htbp]
	\centering
		\includegraphics[scale=0.5]{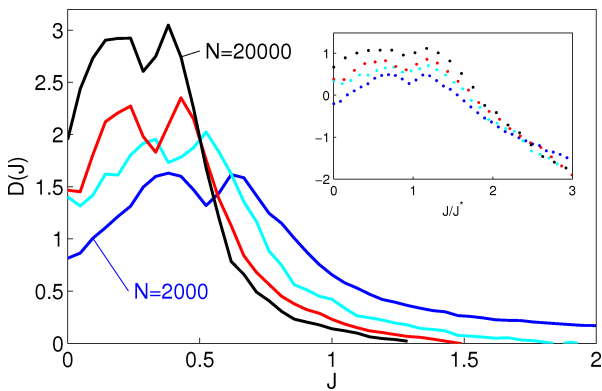}
	\caption{(Color online) Numerically measured diffusion coefficient during the QSS for $N=2.10^3$, $5.10^3$, $10^4$ and $2.10^4$. The inset shows $\log (D)$ with respect to the rescaled action $J/J^{\ast}$ in the vicinity of the resonance.}
\label{fig:D_mesure}
\end{figure}

Starting from an initial WB distribution at an energy $U=0.69$ (as shown in Fig.~\ref{fig:evo_M}(b)), we ran several realizations of the system. For each one, a time interval is selected so that the QSS is well established, with the mean-field $M(t)$ oscillating about a constant $M_{0}$ value. Accordingly with \cite{CampaPRE}, we found $M_{0}$ to scale as $N^{-\frac{1}{2}}$. Then we let test particles evolve under the recorded $M(t)$. Their phase-space coordinates allow us to calculate the corresponding action using Eq.~(\ref{eqn:Jh}). We analyzed their trajectories as if they were realizations of Langevin dynamics, in which noise arises from the mean field fluctuations $\delta M(t)$, modeled here by a Gaussian process $\xi$ verifying $\left\langle\xi(t)\xi(t^\prime)\right\rangle=C(N)\delta(t-t^\prime)$. Let us note that such a stochastic approach is reminiscent of the one used by Chandrasekhar in seminal works \cite{Chandra1949} on the relaxation of stellar systems as a result of finite-$N$, discreteness, noise. However, noise is not driven here by particle encounters or another binary effect but by a non-local collective effect. The standard deviation $\sqrt{C(N)}$ is found to scale as $N^{-\frac{1}{2}}$ like $M_{0}$. Namely, one has
\begin{equation}
\dfrac{\mathrm{d}J}{\mathrm{d}t}= -\dfrac{\mathrm{d}V_\mathrm{eff}}{\mathrm{d}J} + \sqrt{C(N)D(J)} \xi(t),
\label{eqn:Langevin}
\end{equation}
where the effective potential $V_\mathrm{eff}$ accounts for the collective effects which are not captured by this noise modeling. Practically speaking, we found this potential to be negligible, reducing the model to a non-biased random walk. More precisely, the diffusion properties can be retrieved through the quantity $\langle\delta J^2\rangle$, defined by
\begin{equation}
\left\langle \delta J^{2}\right\rangle(J,\delta t)=\left.
\left\langle \left( J(t+\delta t)-J(t)\right) ^{2}\right\rangle
\right\vert _{J(t)=J},  \label{eqn:getH}
\end{equation}
where the brackets denote an average over particles. The diffusion coefficient $D(J)=(C(N))^{-1} \langle \delta J^2 \rangle/\delta t$ is then obtained using a simple linear fit in the mesoscopic lapse of time corresponding to the diffusive regime. Figure~\ref{fig:D_mesure} shows the resulting local diffusion coefficient $D(J)$ computed for different values of the number of particles $N$.

Note that in spite of the factorization of $C(N)$, the diffusion coefficient still depends on the number of particles. This is not surprising since $D(J)$ implicitly depends on the value of $M_{0}$ during the QSS, which depends on $N$. Figure~\ref{fig:D_mesure} gives a crucial information on the action locality of the HMF dynamics and on the role of the number of degrees of freedom on these locality properties within the $M=o(1)$ QSS regime.

\begin{figure}[htbp]
	\centering
		\includegraphics[width=0.95\columnwidth]{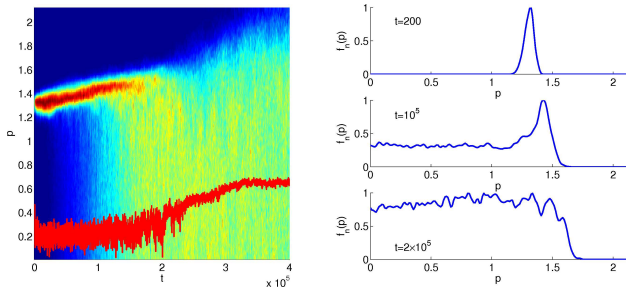}
	\caption{(Color online) (Left) Color plot of the normalized action distribution of test particles starting from a $\delta$-distribution at $p=1.35$, evolving under the fluctuations of $M(t)$ for $N=5000$ particles. The red curve is the instantaneous separatrices mean position, equal to $4\sqrt{M(t)}/\pi$. The right panel displays three snapshots of the velocity distribution. As time increases, the initial peak slowly fades away, allowing particles to explore the whole action space.}
\label{fig:evo_cluster}
\end{figure}

Let us first consider the curve $D(J)$ obtained for a given (large) $N$. Starting from the center of the resonance at $J=0$, $D$ increases and attains its maximal values on both sides of the separatrices action line $ J=J^{\ast}$ with a double humped shape and a local minimum centered on $J^{\ast }$. As the action further increases, $D$ smoothly decreases towards a vanishing value. Comparing now the curves obtained for different $N$, it turns out that the curve pattern just described shifts towards the left as $N$ increases which simply reflects the fact that $J^{\ast}\propto\sqrt{M_0}$ is decreasing with $N$. Moreover, $D$ happens to increase with $N$ in this shrinking action domain swept by the separatrices. On the contrary, for action values far enough from $J^{\ast}$, $D$ is found to decrease with $N$, the crossing between both behaviors being visible on the inset of Fig.~\ref{fig:D_mesure}. All this implies that the underlying diffusive process strongly depends on the action value: when $0 \leq J \lesssim J^{\ast}$, particles diffuse strongly, whereas particles in the action domain $J^{\ast}\ll J\leq J_{\mathrm{max}}(t)\sim 1$ diffuse very slowly. Therefore the timescale of the thermalization process, or equivalently the QSS lifetime, will be controlled by those very slowly diffusing particles hardly escaping from the $J=O(1)$ domain.  This is illustrated on Fig.~\ref{fig:evo_cluster}, which captures the diffusion of particles initially located close to the edge of the QSS distribution\footnote{Note that the QSS distribution, that is reached after some transient stage, is wider in terms of momenta than the initial waterbag $f_{WB}$.}, slowly diffusing towards the separatrices, on a timescale comparable to the QSS one.

Let us now further analyze the locality properties of the transport, interpret the two limiting regimes identified on Fig.~\ref{fig:D_mesure} and propose an explanation for the scaling of the diffusion coefficient in the $J\sim1$ domain that is the key of the scaling of the $M=o(1)$ QSS lifetime. The diffusion equation arising from the phenomenological Langevin equation given by Eq.~(\ref{eqn:Langevin}) may be written as
\begin{equation}
\frac{\partial P}{\partial t}=\frac{C(N)}{2}\dfrac{\partial }{\partial J}\left[ D(J) \dfrac{\partial P}{\partial J}\right]\equiv \dfrac{\partial }{\partial J}\left[ D_{\mathrm{eff}}(J) \dfrac{\partial P}{\partial J}\right],\label{FP}
\end{equation}
in which the effective local diffusion coefficient $D_{\mathrm{eff}}(J)$ is defined. Moreover, there is a single action characteristic scale in the HMF system given by $J^{\ast}$. In order to compare the $M=O(1)$ and $M=o(1)$ QSSs cases, it is convenient to move to dimensionless variables, namely use the rescaled action $\hat{J}=J/J^{\ast}$, time $\hat{t}=J^{\ast}t$ and probability density $\hat{P}=J^{\ast}P$. In terms of these variables, the Fokker-Planck equation (FPE)~(\ref{FP}) becomes
\begin{equation}
\frac{\partial \hat{P}}{\partial \hat{t}}=\frac{C(N)}{2J^{\ast 3}}\frac{%
\partial }{\partial \hat{J}}\left[ D(J^{\ast} \hat{J}) \frac{\partial \hat{P}}{%
\partial \hat{J}}\right].  \label{FP_rescaled}
\end{equation}
For low action values, numerical measurements indicate that $D(J)=D(J^{\ast} \hat{J})$ scales as $J^{\ast -2}\propto N^{1/2}$ in the QSS regime, so that, in this domain the effective diffusion coefficient in Eq.~(\ref{FP_rescaled}) diverges with $N$ as $J^{\ast -1}\propto N^{1/4}$. This is a ``strong'' noise limit compared with the $M=O(1)$ QSS case in which the latter scales as $N^{-1}$~\cite{EttFir2}: Here the (infinitesimal) fraction of quasi-resonant particles diffuses strongly due to the fact that the mean value and fluctuations of the mean field are of the same order.
\begin{figure}[htbp]
\centering
\includegraphics[width=0.95\columnwidth]{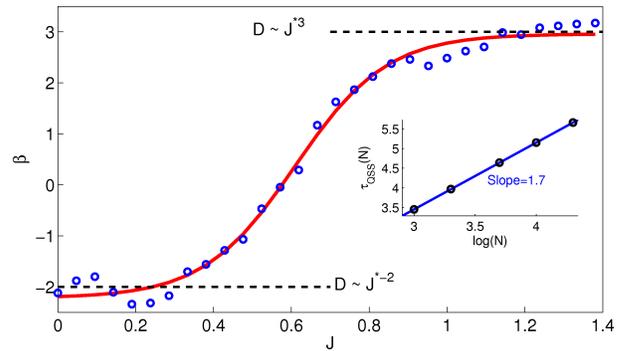}
\caption{At any given $J$, the dependency of $D(J)$ with $J^\ast$ can be well approximated by a power law as in Eq.~(\ref{Ansatz}). The exponent $\beta$ is plotted here with respect to $J$. Each circle point has been obtained from a linear fit of the points $\log(D(J))$ w.r.t. $\log(J^{\ast})$ corresponding to the $N=2.10^3$, $5.10^3$, $10^4$ and $2.10^4$ measurements. The plain curve is the best fitting curve of the hyperbolic tangent form. The inset shows the numerical estimate of the QSS lifetime according to Eq.~(\ref{eqn:tqss}) using the latter fit.}
\label{fig:evo_beta}
\end{figure}

Let us now estimate the scaling of $D$ with $N$ in the macroscopic fraction of the phase space for which $J$ is of order one. The crucial point here is to observe that for a particle with $J \sim 1$ the relative effect on its motion of the mean-field fluctuations, scaling as $N^{-\frac{1}{2}}$ are much weaker than for a particle having $J \sim J^{\ast}$. More precisely, this effect will be of the same order as for a generic particle in the $M=O(1)$ QSS case, having $J \sim 1$,  for which the diffusion rely on the mean-field fluctuations of order $N^{-\frac{1}{2}}$. Therefore one predicts that, in the limit $\hat{J}\gg1$, corresponding to $J \sim 1 \gg J^{\ast}$, the effective diffusion coefficient expressed in the rescaled FPE (\ref{FP_rescaled}) scales as $C(N)$ in the $M=o(1)$ QSS, namely as $N^{-1}$ as it would scale in the $M=O(1)$ QSS case for $\hat{J} \sim 1$ \cite{EttFir2}. In Eq.~(\ref{FP_rescaled}), this yields $D(J)\propto {J^{\ast}}^{3}$ when $J \gg J^\ast$.

In a final step to estimate the QSS lifetime, we made the Ansatz that, for any given $J$, the dependency with $N$ (through ${J^\ast}$) of the diffusion constant could be cast into a simple power-law
\begin{equation}
D(J) \propto {J^{\ast}}^{\beta(J)}. \label{Ansatz}
\end{equation}
Figure~\ref{fig:evo_beta} shows the numerical measurement of $\beta(J)$, hence confirming the asymptotic behavior $\beta(J)\rightarrow 3$ for $J \sim 1 \gg J^{\ast}$ as well as exhibiting the strong noise resonant layer behavior $\beta(J)\rightarrow -2$ for $J\rightarrow 0$.

In the original time and action variables, the QSS lifetime can now be estimated by calculating the time needed by a macroscopic fraction of particles to escape from the hardly-diffusing action domain. This would provide us with the scaling with $N$ of such a QSS, as we did in \cite{EttFir2}. However, one can also obtain the desired scaling by calculating the mean first passage time of such particles through the separatrices, since this represents a characteristic timescale of the phase space visit. Let us remark that the latter time is significantly smaller than the actual QSS lifetime, but holds the same scaling as it comes from the same diffusion equation. The Fokker-Planck operator in Eq.~(\ref{FP}) being self-adjoint, it is possible to analytically calculate \cite{Siegert,Tuckwell} such a mean first passage time $\left\langle\tau\right\rangle$. This yields
\begin{equation}
\left\langle \tau \right\rangle = \int^{J_{0}}_{J^{\ast}} \frac{2J}{C(N)D(J)} \mathrm{d}J \underset{N\rightarrow \infty}{\sim} \frac{N^{1+\beta(J_0)/4}}{\sqrt{\left|\beta^{\prime\prime}(J_0)\right|}},
\label{eqn:tqss}
\end{equation}
where we have assumed that $\beta^{\prime\prime}(J_0)\neq 0$. By simply replacing $\beta$ by its asymptotic value far from the separatrices, one finds $\left\langle \tau \right\rangle\propto N^{7/4}$. This scaling falls in the range $1.7\pm0.1$ already measured for QSSs with $N$ in the range $[2.10^{3};2.10^{4}]$ ~\cite{Yamaguchi,Zanette,CampaPRE}. Using a hyperbolic tangent numerical fit for $\beta$, which is plotted on Fig.~\ref{fig:evo_beta}, and taking $J_{0}$=1.5, one recovers for the QSS scaling the famous $N^{1.7}$ scaling law in the same $N$ range.

Finally, for some upper critical energy density $U>U_{c}$, the QSS timescale was measured to diverge as $\mathrm{e}^{N}$ \cite{CampaPRE}. We conjecture that this scaling is related to the existence of some KAM tori preventing diffusion through some action threshold in the effective hamiltonian dynamics (\ref{pendulum}) and to the fact that some finite fraction of the phase space behaves in an almost regular manner. The $\mathrm{e}^{N}$ QSS timescale should then rely on a residual, much slower, diffusion of the Arnold type \cite{Arnold}, that requires more than two degrees of freedom, expressing the fact that, due to the self-consistency, the HMF model is indeed a $N$ degrees of freedom system. 

The method to estimate QSS timescales presented here is expected to be applicable to a wide range of systems in which a Fokker-Planck description is meaningful. It relies on two essential ingredients: the evaluation of the diffusion coefficient in the less diffusive QSS phase space domain and on the use of a timescale of phase space visit, e.g. a mean first passage time.

\begin{acknowledgments}
WE wishes to thank R.~Vasseur, A.~Lazarescu and C.~Sire for useful discussions and acknowledges funding from Ecole Polytechnique through a Gaspard Monge Scholarship. MCF thanks Y.~Kominis for a useful communication. Assistance on parallel computing by A.~F.~Lifschitz is gratefully acknowledged.
\end{acknowledgments}

\end{document}